\documentclass{aa}
\usepackage{graphicx}
\def\gtrsim{\mathrel{\hbox{\rlap{\hbox{\lower4pt\hbox{$\sim$}}}\hbox{$>$}}}}
\def\ltsim{\mathrel{\hbox{\rlap{\hbox{\lower4pt\hbox{$\sim$}}}\hbox{$<$}}}}

\def\bz{\hbox{$\langle B_z\rangle$}}
\begin{document}
\title{Discovery of the pre-main sequence progenitors\\of the magnetic Ap/Bp stars?\thanks{Based on observations obtained at the Canada-France-Hawaii Telescope (CFHT) which is operated by the National Research Council of Canada, the Institut National des Sciences de l'Univers of the Centre National de la Recherche Scientifique of France, and the University of Hawaii, as well as on observations obtained at the European Southern Observatory VLT during observing runs 072.C-0442, 272.C-5063 and 074.C-0447.}}


   \author{G.A. Wade\inst{1}, D. Drouin\inst{1}, S. Bagnulo\inst{2}, J.D. Landstreet\inst{3}, E. Mason\inst{2}, J. Silvester\inst{1,4}, E. Alecian\inst{5},\\ T. B\"ohm\inst{6}, J.-C. Bouret\inst{7}, C. Catala\inst{5}, J.-F. Donati\inst{6}}

   \offprints{G.A. Wade, {\tt Gregg.Wade@rmc.ca}}
   \institute{Dept. of Physics, Royal Military College of Canada, 
   PO Box 17000, Stn Forces, Kingston, Canada K7K 4B4
\and
   European Southern Observatory, Casilla 19001, Santiago 19, Chile
\and
   Dept. of Physics \& Astronomy, University of Western Ontario, London, Canada, N6A 3K7
\and
   Department of Physics, Queen's University, Kingston, Canada 
\and
   Obs. de Paris LESIA, 5 place Jules Janssen, 92195 Meudon Cedex, France 
\and
   Obs. Midi-Pyr\'en\'ees, 14 Avenue Edouard Belin, Toulouse, France
\and
   Laboratoire d'Astrophysique de Marseille, Traverse du Siphon - BP 8, 13376 Marseille Cedex 12, France}
             
   \date{Received ??; accepted ??}

   \abstract{We report the discovery, using FORS1 at the ESO-VLT and ESPaDOnS at the CFHT, of magnetic fields in the young A-type stars HD 101412, V380 Ori and HD 72106A. Two of these stars (HD 101412 and V380 Ori) are pre-main sequence Herbig Ae/Be (HAeBe) stars, while one (HD 72106A) is physically associated with a HAeBe star. Remarkably, evidence of surface abundance spots is detected in the spectra of HD 72106A. The magnetic fields of these objects display intensities of order 1 kG at the photospheric level, are ordered on global scales, and appear in approximately 10\% of the stars studied. Based on these properties, the detected stars may well represent pre-main sequence progenitors of the magnetic Ap/Bp stars. The low masses inferred for these objects (2.6, 2.8 and 2.4~$M_\odot$ represent additional contradictions to the hypothesis of Hubrig et al. (2000), who claim that magnetic fields appear in intermediate-mass stars only after 30\% of their main sequence evolution is complete. Finally, we fail to confirm claims by Hubrig et al. (2004) of magnetic fields in the Herbig Ae star HD 139614.}

\titlerunning{Discovery of the pre-main sequence progenitors of the magnetic Ap/Bp stars?}
\authorrunning{Wade et al.}

   \maketitle

\section{Introduction}
 
Approximately 5\% of main sequence stars of intermediate mass and spectral types A and B display strong ($\sim$kG), globally-ordered magnetic fields (the Ap/Bp stars). These magnetic fields have been frequently suggested (e.g. Mestel 2001, Moss 2001) to be the magnified remnants of interstellar magnetic field swept up during the process of star formation (the primordial fossil field hypothesis). If the magnetic fields of Ap/Bp stars are indeed primordial fossils, we should expect to observe a similar fraction of pre-main sequence (PMS) stars of intermediate mass which host fossil fields of similar structure to the Ap/Bp stars, and similar or somewhat weaker strength.

In this Letter we present first results from two independent surveys aimed at characterising magnetic fields in pre-main sequence A and B type stars. 

   \begin{figure*}[t]
   \centering
   \includegraphics[width=5.9cm]{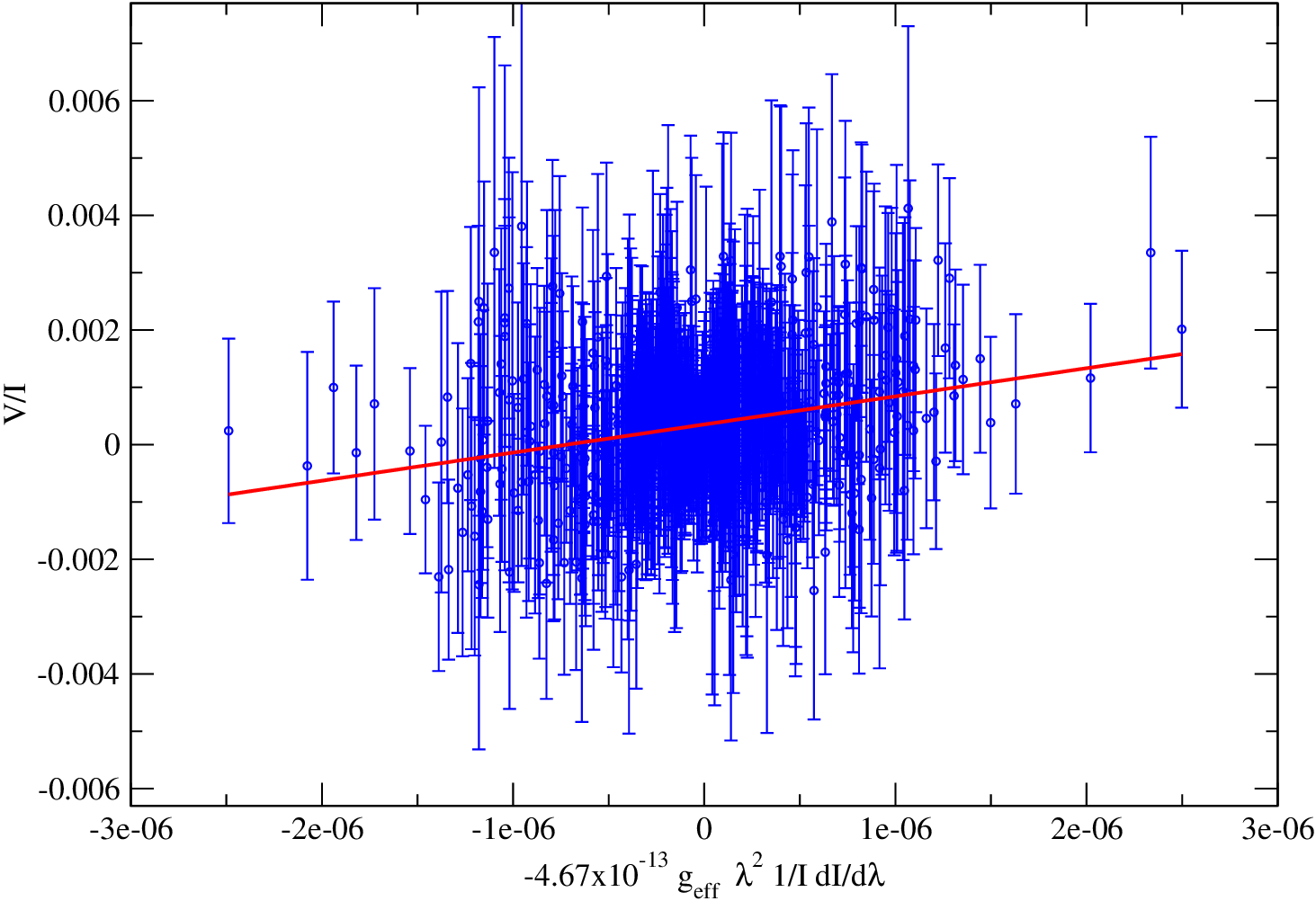}\hspace{1mm}\includegraphics[width=5.9cm]{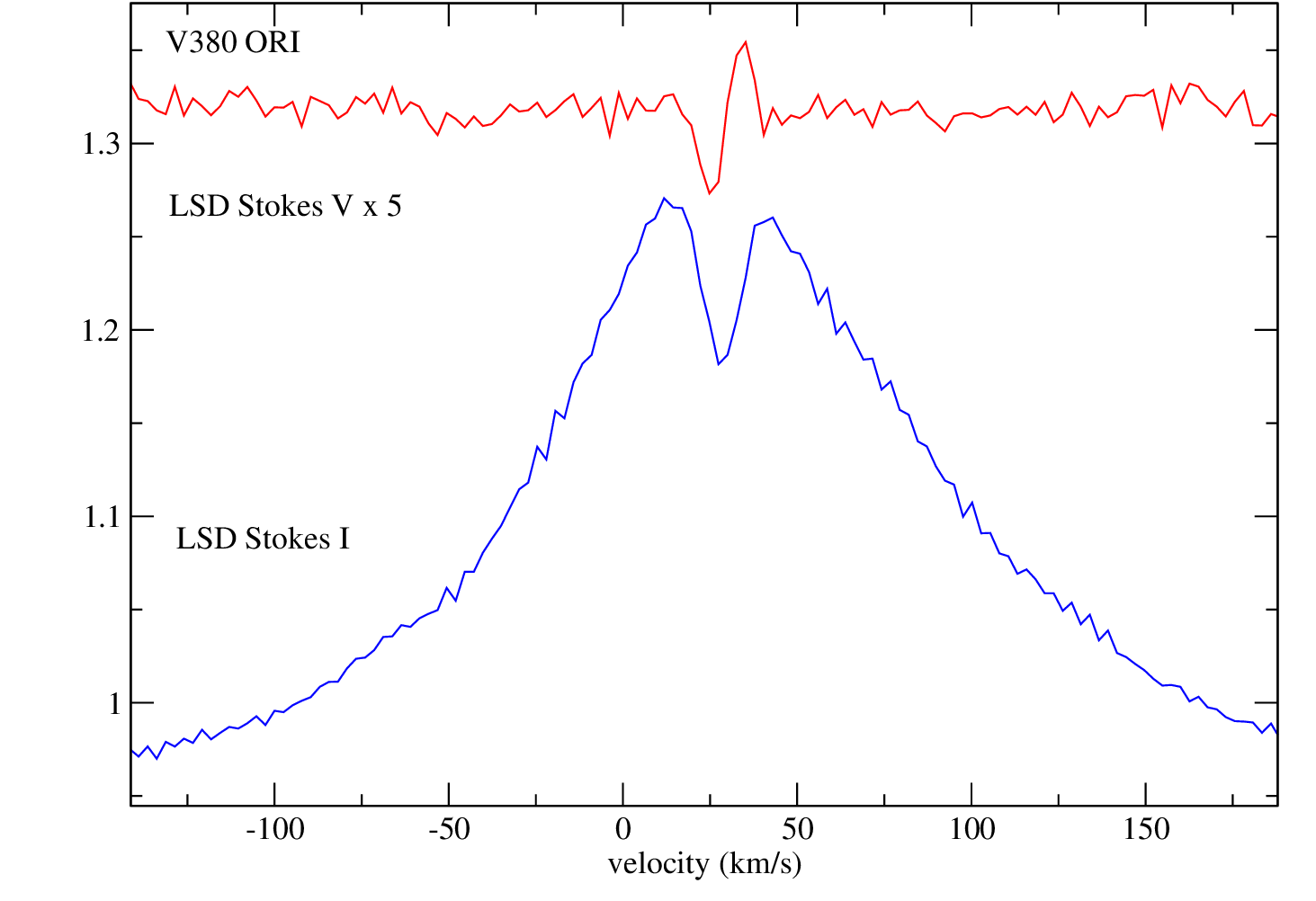}\hspace{1mm}\includegraphics[width=5.9cm]{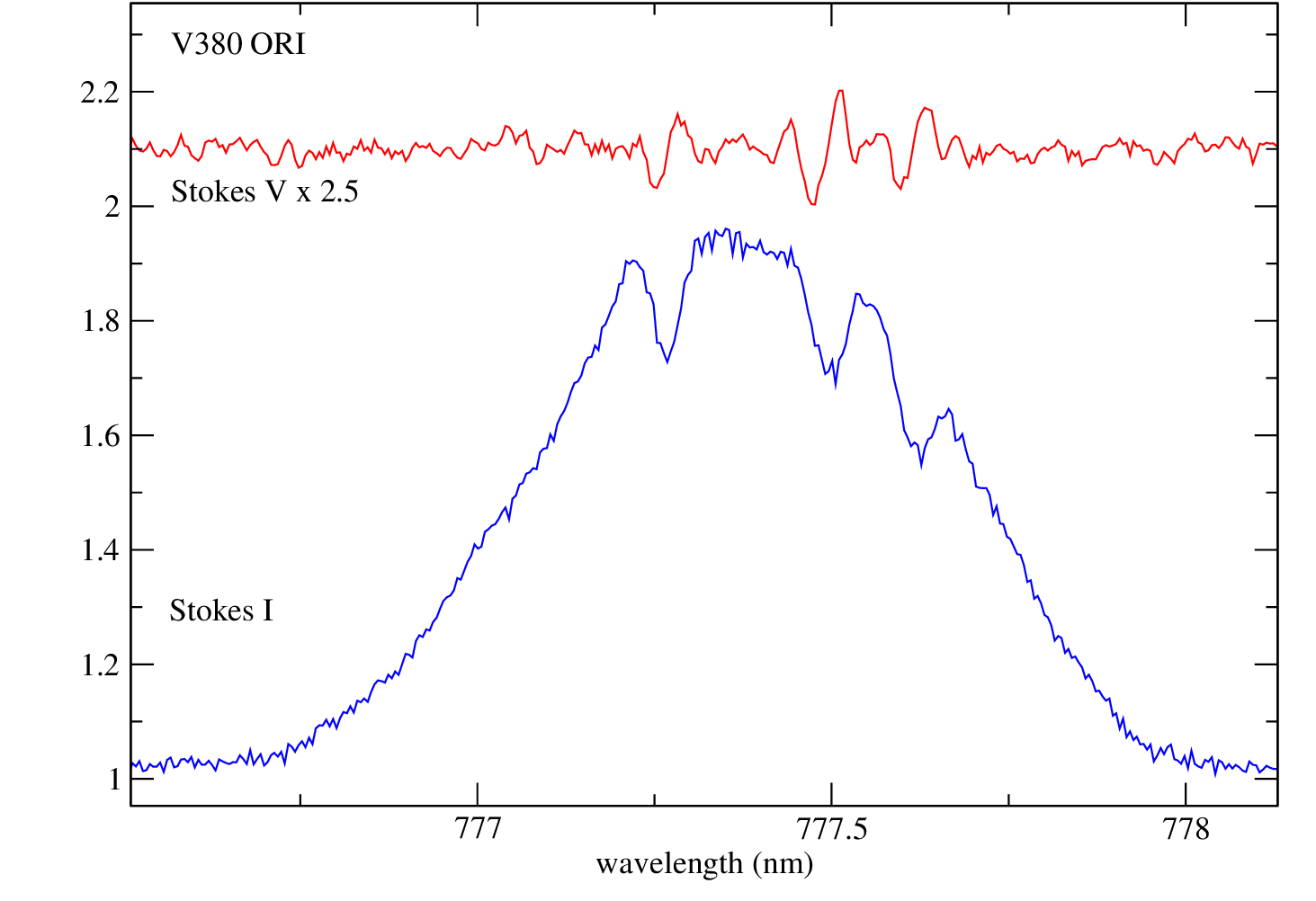}
         \label{}
\caption{Magnetic field diagnoses of:\ {\em Left frame -}\ HD 101412 (FORS1 Balmer-line regression)\ {\em Centre frame -}\ V380 Ori (ESPaDOnS LSD Stokes $I$ and $V$ profiles\ {\em Right frame -}\ V380 Ori (O~{\sc i} 777 nm profiles). The Stokes $V$ signatures detected in the spectrum of V380 Ori correspond to an approximately dipolar surface magnetic field of intensity $\sim 1.5$~kG.}
   \end{figure*}


\section{Observations}

The first survey has been undertaken using the ESO-VLT and the FORS1 spectropolarimeter. High-precision, low-resolution circular polarisation spectra were obtained for 49 PMS Herbig Ae/Be (HAeBe) stars and various polarimetric standards, and the longitudinal magnetic field was inferred from both Balmer lines and metallic lines. The analysis of these data follows the procedure described by Bagnulo et al. (2002) and is described in detail by Drouin (2005).

The second survey was undertaken using the new ESPaDOnS spectropolarimeter (Donati et al. 2005, in prep.)\footnote{see also {\tt www.cfht.hawaii.edu/Instruments}} at the CFHT. High-precision, high-resolution circular polarisation spectra of 20 PMS stars and various polarimetric standards were obtained. Using the Least-Squares Deconvolution (LSD) multi-line analysis procedure (Donati et al. 1997), mean Stokes $I$ and $V$ profiles have been extracted and longitudinal magnetic fields inferred. The analysis of these data is fundamentally similar to that discussed by Shorlin et al. (2002), and will be described in detail by Wade et al. (in preparation).


The log of observations discussed in this paper is reported in Table 1. In the following sections, we describe results for four stars of particular interest.

\begin{table}
\begin{center}
\begin{tabular}{ccccr}
\hline
\noalign{\smallskip}
HJD     & Telescope & Object  & S/N &$\langle B_z\rangle$\\
-2450000     &  & name  & pix$^{-1}$ & (G)\\
\noalign{\smallskip}
\hline
\noalign{\smallskip}
2904.039& VLT& HD 139614   & 2400 &    $-150\pm 50$\\
3332.290& VLT & HD 72106A  & 1525 &   $+195\pm 45$\\
3332.290& VLT & HD 72106B  & 1350 &   $+65\pm 55$\\
3062.297& VLT & HD 101412  & 1135 &   $+430\pm 75$\\
3062.297& VLT & V380 Ori   & 625  &    $+29\pm 35$\\
3422.973& CFHT& HD 72106AB & 250 &  $-15\pm 55$\\
3423.925& CFHT& HD 72106AB & 225 &  $-125\pm 60$\\
3423.925& CFHT& V380 Ori   & 125 &    $-460\pm 70$\\
3423.076& CFHT& HD 139614  & 270 &   $-20\pm 25$\\
3424.096& CFHT& HD 139614  & 260 &   $-35\pm 25$\\
\hline\hline\noalign{\smallskip}
\end{tabular}
\caption[]{Journal of circular polarisation observations of the stars discussed in this Letter.}
\label{tab:journal}
\end{center}
\end{table}

\section{HD 101412}

HD 101412 (PDS 057) is a HAeBe star (Th\'e et al. 1994) associated with the Centaurus and Crux star forming regions (SFRs). Vieira et al. (2003) derive an effective temperature from analysis of UBVRI photometry of 9500 K. No parallax data is available for this star, although based on the reported distance to the DC 295.0+1.3 SFR within the Centaurus/Crux complex with which HD 101412 is associated (Vieira et al. 2003; at a distance of 500-700 pc; Corradi et al. 1997), we can estimate its position on the HR diagram. This position, interpreted using the model evolutionary tracks of Palla \& Stahler (1993), indicates that HD 101412 has a mass of $2.6\pm 0.3~M_\odot$, and appeared at the birthline only about 2 Myr ago.

A longitudinal magnetic field of $\bz=+430\pm 75$~G was detected in the FORS1 measurement, in both Balmer lines and metallic lines. The field diagnosis using the Balmer-line regression method is illustrated in Fig. 1 (left panel).

   \begin{figure*}[t]
   \centering
   \includegraphics[width=5.9cm]{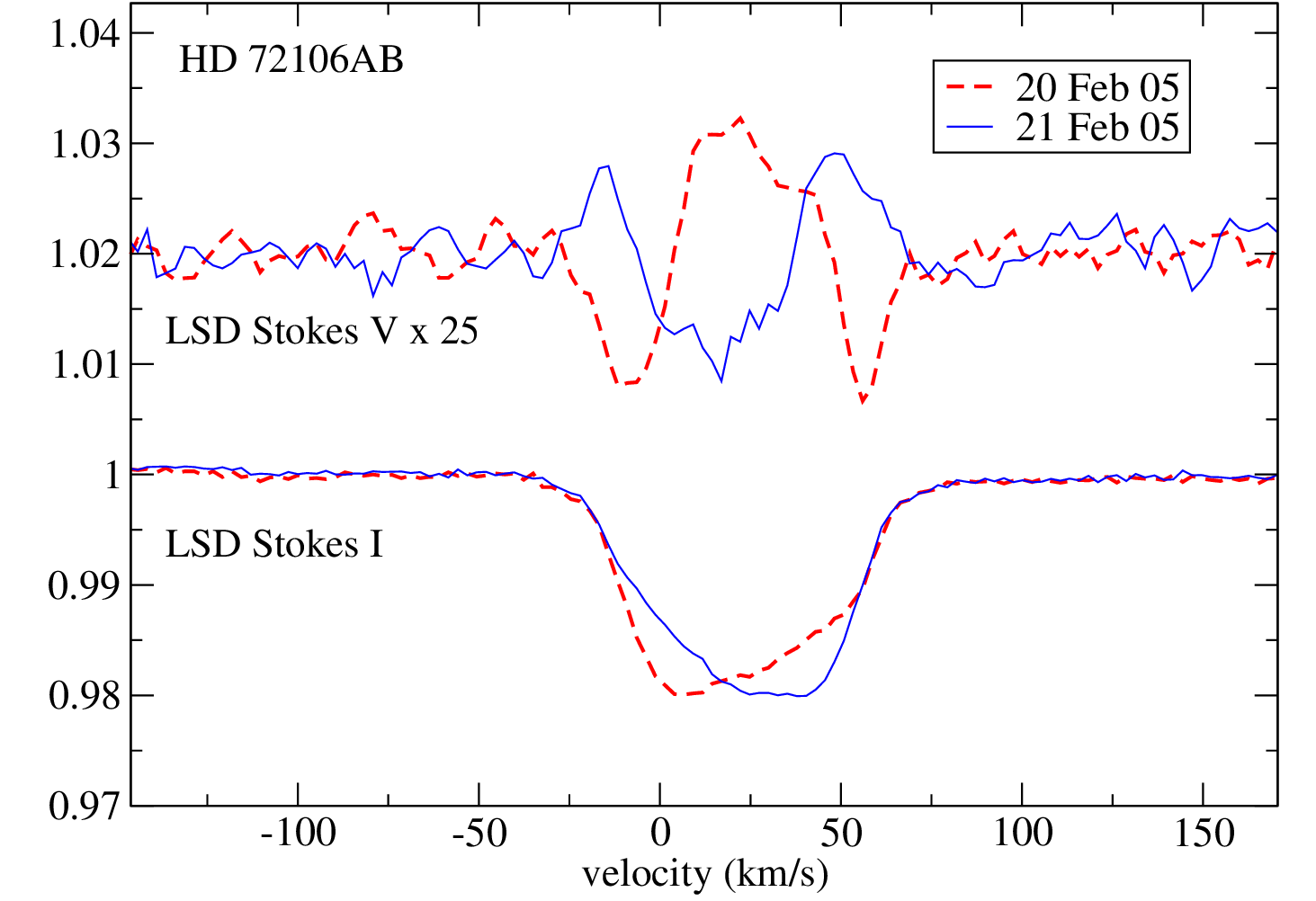}\hspace{1mm}\includegraphics[width=5.9cm]{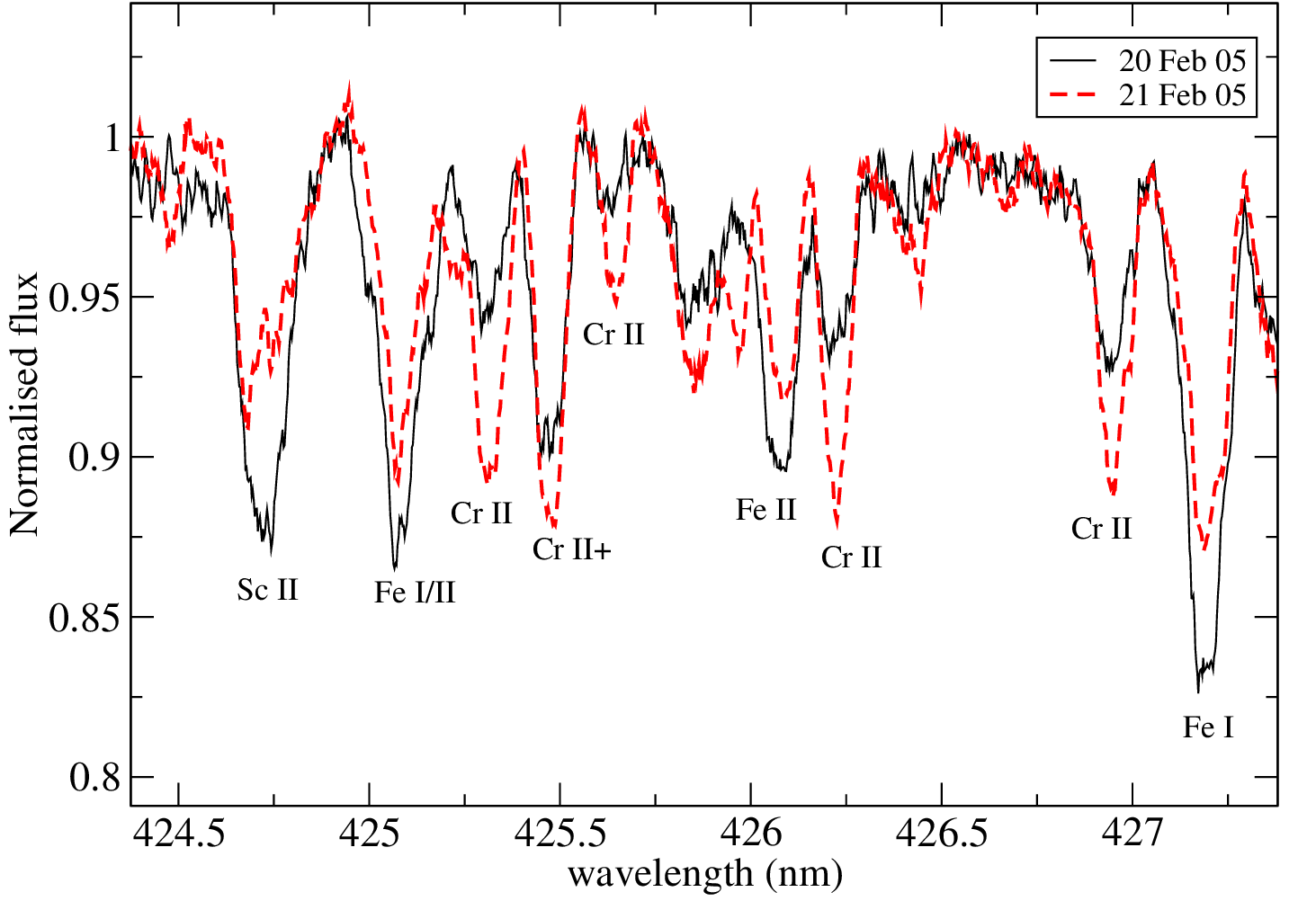}\hspace{1mm}\includegraphics[width=5.9cm]{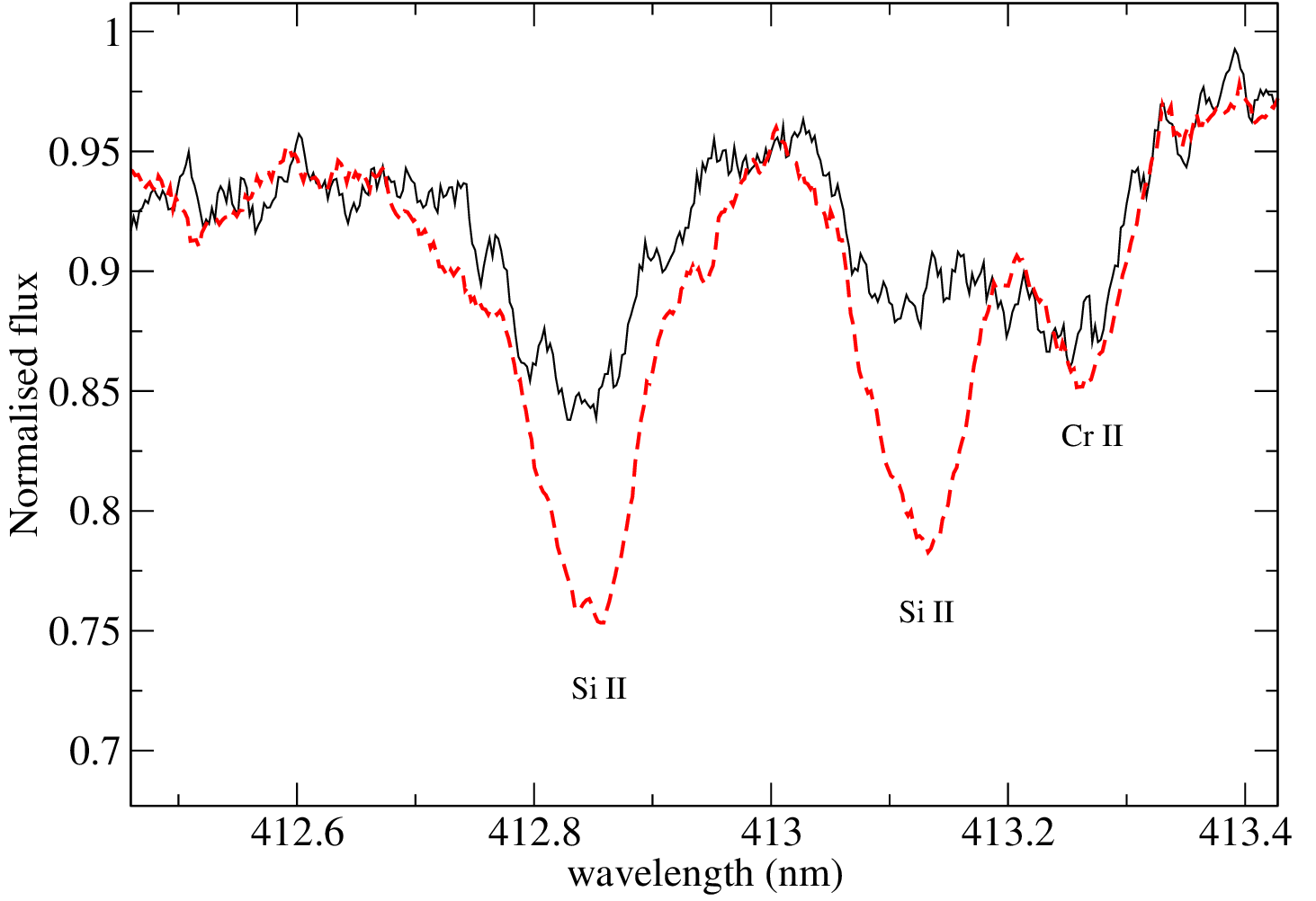}
         \label{}
\caption{{\em Left frame -}\ Magnetic field diagnosis of HD 72106 (ESPaDOnS LSD Stokes $I$ and $V$ profiles). The Stokes $V$ signatures detected in the spectrum of HD 72106 correspond to an approximately dipolar surface magnetic field of intensity 1 kG.\ {\em Left frame -}\ Line profile variations of O~{\sc i}, Si~{\sc ii}, Sc~{\sc ii}, Cr~{\sc ii} and Fe~{\sc ii} observed in the two ESPaDOnS spectra of HD 72106 ({\em solid black line}\ - 20 Feb 05, {\em dashed red line}\ - 21 Feb 05)}
   \end{figure*}

\section{V380 Ori}

V380 Ori is a HAeBe star (Hillenbrand et al. 1992), and a member of the Orion Complex. Hillenbrand et al. (1992) report the effective temperature and luminosity of this object (10700 K and 85~$L_\odot$ respectively) from which we derive an approximate mass of $2.8\pm 0.3~M_\odot$, and an age of about 1 Myr. 

The longitudinal magnetic field of V380 Ori measured with FORS1, using both Balmer lines and metallic lines, was $\bz=+29\pm 35$~G. The ESPaDOnS LSD Stokes $I$ profile (extracted using a solar abundance, 10000 K line mask) shows a wide emission skirt with a central, apparently photospheric sharp central absorption core. Associated with this core, a strong Stokes $V$ signature is detected, corresponding to an longitudinal field of $\bz=-460\pm 70$~G. Stokes $V$ signatures, of similar shape, are also observed in some individual lines in the reduced spectrum. The LSD profiles, along with the Stokes $V$ signatures detected in the three lines of the O~{\sc i} 777 nm triplet, are shown in Fig. 1 (centre and right panels). Due to the strong emission in the mean metal line profile, it is natural that this relatively strong field was not detected at low resolution with FORS1.

\section{HD 72106}


HD 72106 (PDS 031) is a visual double with a separation $\rho=0.8$ arcsec and a magnitude difference $\Delta m_V=0.8$ (Harkopf et al. 1996, ESA 1997). It is associated with the Vela SFR and the Gum Nebula (Veira et al. 2003), at a distance of $288^{+202}_{-84}$ pc according to the Hipparcos parallax. Based on the $B-V$ colours (Fabricius \& Makarov 2000) both components are A-type stars, while Vieira et al. (2003) report that the fainter secondary component exhibits H$\alpha$ emission and significant infrared excess, and characteristics of a relatively evolved HAeBe star. It appears, based on the large proper motions and lack of any relative motion of the components(Harkopf et al. 1996), that they are physically associated.

Magnetic field was detected with FORS1 using both Balmer lines and metallic lines, and subsequently confirmed with ESPaDOnS. The Least-Squares Deconvolved Stokes $I$ and $V$ profiles of HD 72106 obtained using ESPaDOnS are illustrated in Fig. 2 (left panel). The magnetic field is detected in the brighter, hotter primary component of the system\footnote{This is established based on the FORS1 and ESPaDOnS field detections, supplemented by the combined and individual Stokes $I$ spectra}, whereas H$\alpha$ emission (which we also detect) is observed to originate from the cooler, fainter secondary component (Vieira et al. 2003, this work). The positions of the HD 72106 components on the HR diagram (using temperatures derived from fitting the FORS1 Balmer line profiles with ATLAS9 models and luminosities inferred from the parallax distance) are fully consistent with this system being at the end stages of its pre-main sequence evolution, with an age of $10$~Myr since emergence at the birthline. The primary component has an approximate mass of $2.4\pm 0.4~M_\odot$, while the secondary has a mass of $1.75\pm 0.25~M_\odot$.  

The longitudinal field detected in the primary component of HD 72106 by FORS1 is $\langle B_z\rangle=+195\pm 40$~G. The ESPaDOnS measurements both correspond to significant levels of circular polarisation within the mean line profile and marginal levels within individual profiles of strong, magnetically-sensitive lines (such as Fe~{\sc ii} 501.8 nm). As is evident in Fig. 2 (left frame), the two LSD Stokes $V$ signatures are crossover signatures of opposite sign, implying that the large-scale topology of the visible field has changed significantly during $\sim 24$ hours (and therefore that the field is structured on global scales). The measured longitudinal field intensity, the inferred large-scale structure of the surface field, and the variability of the field (whose timescale is consistent with a rotational period of approximately 2 days, as derived from the observed variability, the inferred radius and measured $v\sin i$ of 45~{km/s}) all strongly suggest that the magnetic field of HD 72106 is similar to those of main sequence Ap stars. The lack of detection of a magnetic field ($\langle B_z\rangle=+65\pm 55$~G) in HD 72106B, if confirmed, could also have important implications for field origin scenarios.

Remarkably, clear element-dependent line profile varability is observed in the ESPaDOnS spectra of HD 72106. As illustrated in Fig. 2 (centre and right panels), lines of Fe, Cr, Sc and Si (as well as O, Mg, and Ca) are all clearly variable. The variability phasing differs from element to element, strongly suggesting the presence of abundance spots on the stellar surface, analogous to the Ap/Bp stars. 



\section{HD 139614}

HD 139614 (PDS 395) is a HAeBe star with a relatively cool temperature (7600 K, Vieira et al. 2003; 8000 K, Acke \& Waelkens 2004). This star is associated with the Lupus/Ophiucus complex, in the Lupus SFR at a distance of 110-160 pc (Franco 1990). The HR diagram position (established assuming the SFR distance) implies that HD 139614 is relatively evolved, and has a mass of $1.6\pm 0.2~M_\odot$. 

Hubrig et al. (2004) reported the detection of a longitudinal magnetic field of intensity $\langle B_z\rangle=-450\pm 93$~G based on analysis of polarised spectra obtained with FORS1, using the Balmer-line method of Bagnulo et al. (2002). As reported by Drouin (2005), a re-analysis of the (publically-available) data obtained by Hubrig et al. confirms an apparent field detection in Balmer lines ($\langle B_z\rangle=-320\pm 75$~G, but curiously provides no detection using the entire spectrum (Balmer lines and metallic lines: $\langle B_z\rangle=-73\pm 40$~G). 

In order to further test for the presence a magnetic field in HD 139614, we have obtained two circularly polarised spectra of this star using ESPaDOnS. The observations are characterised by a rich, non-variable metallic-line spectrum broadened slightly by rotation (ideally suited for LSD). We find no evidence for any circular polarisation in individual lines in the high-S/N (250:1) spectra. Using LSD (with a solar abundance 8000 K line mask) we detect no circular polarisation within the mean line, and constrain the longitudinal magnetic field $\langle B_z\rangle$ to be weaker than 75~G (3$\sigma$ upper limit) on both nights. We can rule out a longitudinal field at the level claimed by Hubrig et al. (2004) at 18$\sigma$ confidence (on both nights). For illustration, we overplot in Fig. 3 the lowest-amplitude calculated Stokes $V$ signature, corresponding to the mean profile, required to produce a longitudinal field of $-450$~G. The presence of such a signature can also be ruled out with 18$\sigma$ confidence. 

\section{Discussion}

The discoveries of magnetic fields in HD 101412 and V380 Ori represent highly significant new results. The inferred positions of these stars on the HR diagram indicate that they have only recently appeared at the birthline (ages of $\sim 1$~Myr). These stars are situated well away from the main sequence, and according to models of their PMS evolution (Palla \& Stahler 1993), deuterium is burning in a radiative mantle covering the outer 10\% of the stellar mass, and the star has yet to thermally re-adjust after the termination of accretion. These stars are at a significantly different evolutionary state, and exhibit significantly different internal structure, than the main sequence Ap/Bp stars. {\em This is a qualitatively new result, and the magnetic field detections require immediate confirmation.}

We have also discovered and confirmed the existence of a strong, globally-structured magnetic field in HD 72106A. We have also established that many elements are distributed inhomogeneously over the stellar surface. HD 72106A is therefore clearly an Ap/Bp star. But is it a pre-main sequence star? According to Vieira et al. (2003), HD 72106B exhibits the characteristics of a HAeBe star. Due to its somewhat larger mass, HD 72106A should be somewhat more evolved, and it is impossible to confidently conclude (given the inferred 10 Myr age of the system) that the primary has not yet reached the main sequence. It is clear however that both HD 72106A and HD 72106B are pre-main sequence or very young main sequence objects (based on their inferred HR diagram positions, the presence of dust and gas in the immediate vicinity of HD 72106B, and their association with the Vela/Gum Nebula SFR).

   \begin{figure}[t]
   \centering
   \includegraphics[width=5.9 cm]{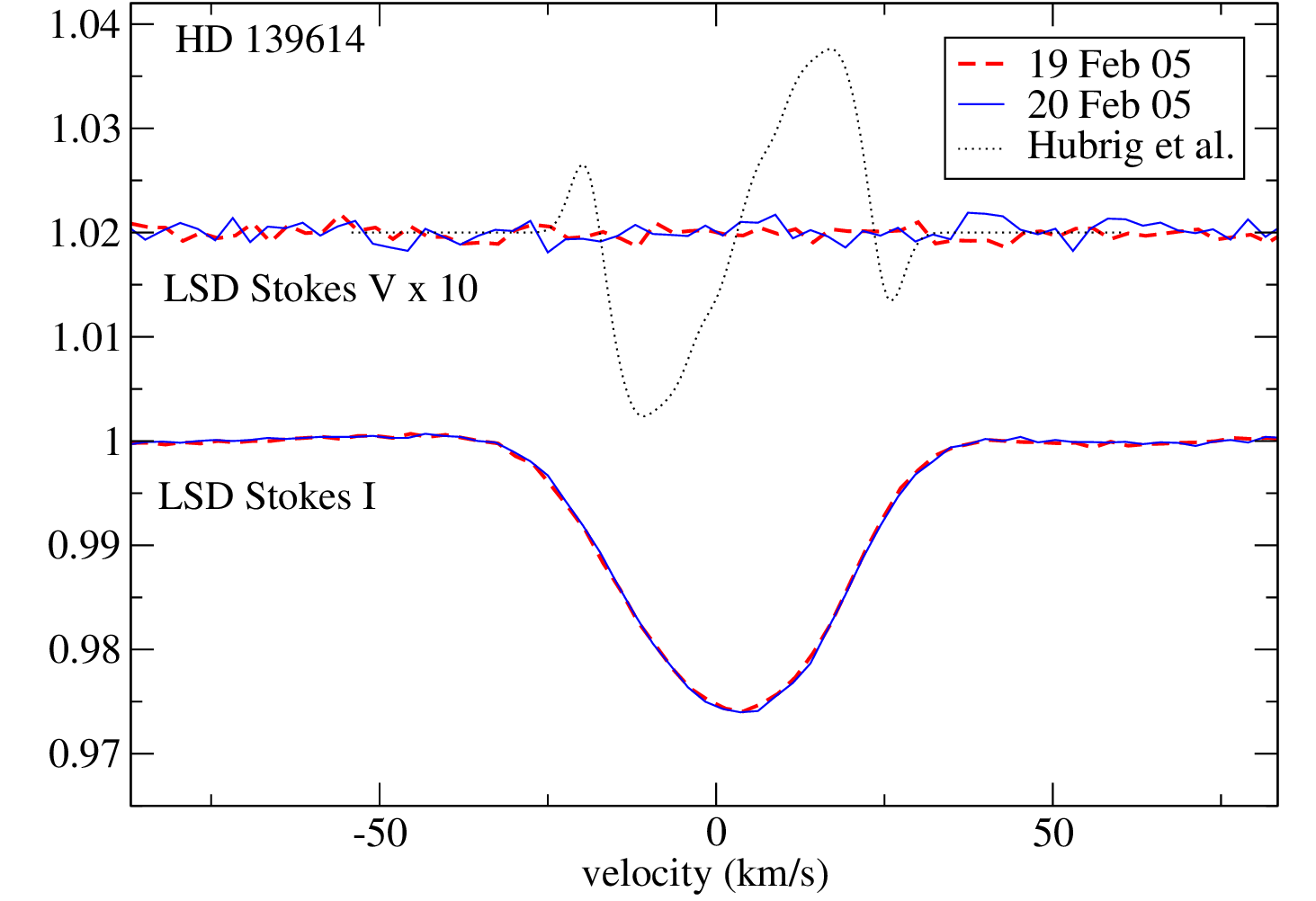}
         \label{}
\caption{Magnetic field diagnoses HD 139614 (ESPaDOnS LSD Stokes $I$ and $V$ profiles). No evidence for magnetic field is found for HD 139614. The dotted curve shows the weakest calculated Stokes $V$ signature corresponding to a -450 G longitudinal field. This signature is ruled out with 18$\sigma$ confidence.}
   \end{figure}

HD 101412, V380 Ori and HD 72106A are inferred by us to have low masses (below $3~M_\odot$). If the masses are confirmed for these very young objects, all would represent clear contradictions to the conclusion of Hubrig et al. (2000), who hypothesise that Ap stars with masses below 3~$M_\odot$ do not become magnetic until completing $\sim 30\%$ of their main sequence evolution. 

The detection of five stars\footnote{The three stars discussed here, as well as the HAeBe stars HD 190073 and HD 200775, reported by Catala et al. (in preparation).} hosting apparently Ap-type magnetic fields in a sample of approximately 50 PMS stars yields an observed incidence of 10\%, which is in global agreement with the incidence prediction performed for this sample by Drouin (2005) assuming the primordial fossil hypothesis. This suggests that the progenitors of the magnetic Ap/Bp stars can probably be found amongst the PMS stars of intermediate mass, with the techniques of magnetic field investigation used in this paper. On the other hand, much of the literature discussion of the activity of HAeBe stars has focussed on dynamo-type field phenomena, and we find no direct evidence of the presence of such fields in the stars studied. A more complete assay (both a broader survey and more detailed investigation of individual objects) of the magnetic characteristics of PMS intermediate mass stars is clearly necessary before more significant conclusions can be drawn.

\begin{acknowledgements}

GAW and JDL acknowledge Discovery Grant support from the Natural Sciences and Engineering Research Council of Canada. 

The ESPaDOnS data were reduced using the data reduction software Libre-ESpRIT (Donati et al. 2005, in prep.), written by J.-F. Donati from Observatoire Midi-Pyr\'en\'ees and made available to observers at CFHT.
\end{acknowledgements}
\end{document}